\documentclass[traditabstract]{aa}
\usepackage{graphicx}
\usepackage{txfonts}
\usepackage{natbib}

\begin{document}

\title{Stellar wobble caused by a nearby binary system: eccentric and inclined orbits}

   \titlerunning{Stellar wobble caused by a nearby binary system}

\author{M.H.M. Morais
          \inst{1}
          \and
          A.C.M. Correia\inst{1,2}}

   \offprints{M.H.M. Morais}

   \institute{Department of Physics, I3N, University of Aveiro,
	     Campus Universit\'ario de Santiago, 3810-193 Aveiro, Portugal \\
             \email{helena.morais@ua.pt}
		\and
           Astronomie et Syst\`emes Dynamiques, IMCCE-CNRS UMR 8028, 
             77 Avenue Denfert-Rochereau, 75014 Paris, France
             }

 \abstract{Most extrasolar planets currently known were discovered by means of an indirect method that measures the stellar wobble caused by the planet.  We previously studied a triple system composed of a star and a nearby binary on circular coplanar orbits. We showed that although the effect of the binary on the star can be differentiated from the stellar wobble caused by a planet, because of observational limitations the two effects may often remain indistinguishable. Here, we develop a model that applies to eccentric and inclined orbits. We show that the binary's effect is more likely to be mistaken by planet(s) in the case of coplanar motion observed equator-on. Moreover, when the orbits are eccentric, the magnitude of the binary's effect may be larger than in the circular case.
Additionally, an eccentric binary can mimic two planets with orbital periods in the ratio 2/1. However, when the star's orbit around the binary's center of mass has a high eccentricity and a reasonably well-constrained period, it should be easier to distinguish the binary's effect from a planet. 
}

  \keywords{stars: planetary systems --
                astrometry --
                celestial mechanics}

\maketitle


\section{Introduction} 

In \citet{Morais&Correia2008}, we studied a triple system consisting of a star perturbed by a binary system.
Our aim was to derive an expression for the star's radial velocity in order to determine whether the binary's effect could be mistaken for that of a planet companion to the star. This question is relevant if one or even both binary components are unresolved
which can happen for instance when these are faint but very common M stars. About $76\%$ of nearby main sequence stars are M-type \citep{Starrysky2001}. Moreover, in the solar neighborhood, over $50\%$ of G and K-stars
\citep{Duquennoy&Mayor1991,Halbwachs_etal2003,Eggenberger_etal2004} and about $30\%$ of M-stars \citep{Fischer&Marcy1992,Delfosse_etal2004} belong to binary or even multiple systems.  

We previously showed that the binary system's effect on the star is a sum of two periodic signals with very close frequencies. These signals, if detected, could be associated with two planets on circular orbits around the star. However, these planets would have very close orbits and we should be able to discard them as unstable systems. Moreover, we derived expressions for the frequencies and amplitudes of these signals, thus were able to identify the hidden binary's parameters.
As explained in \citet{Morais&Correia2008}, our results do not agree with previous work by \citet{Schneider&Cabrera2006} who studied the case of an equal-mass binary and concluded that its effect is a single periodic signal that mimics a planet on an eccentric orbit. 

Nevertheless, in \citet{Morais&Correia2008} we saw that there are realistic situations where we can identify one of the periodic signals but not the other. In this case, we may mistake the binary's effect for that of a planet on a circular orbit around the star. However, we can still apply our theory to compute the hidden binary's parameters and then try to detect these objects. 

Our work was based on a full three-body model but assumed coplanar motion and initial circular orbits for the star and binary.
However, general triple-star systems are likely to have non-coplanar motion and eccentric orbits. Therefore, in this article, we extend our study to the case of three-dimensional non-circular motion of the star and binary system. In Sect.~2, we present the model and in Sect.~3 we discuss the circumstances that lead to the binary being mistaken by one or more planets. In Sect.~4, we compare the theoretical predictions with results obtained from simulations of hypothetical triple systems, we discuss the triple system HD 188753, and we (re)analyze the exoplanets discovered within binary systems. Finally, in Sect.~5 we present our conclusions.


\section{Modeling a star perturbed by a binary system}

\subsection{Equations of motion and radial velocity}

\begin{figure}
  \centering
    \includegraphics[width=8cm]{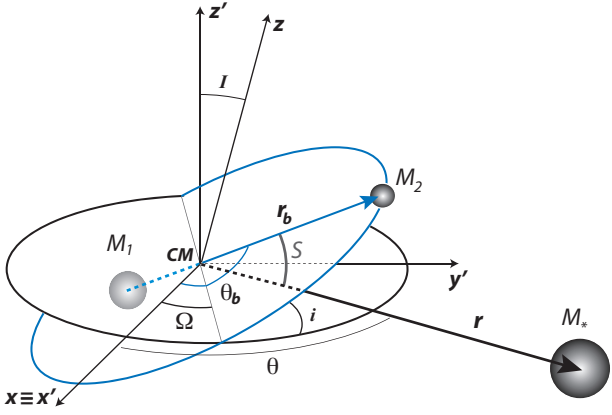}
  \caption{A star with mass $M_{\star}$ is perturbed by a binary system
  with masses $M_{1}$ and $M_{2}$. The Jacobi coordinates $\vec{r}_b$ and $\vec{r}$ are, respectively,
  the inter-binary distance and the distance of  $M_{\star}$ to the binary's center of mass.
  The unperturbed Keplerian described by $\vec{r}$ is in the plane $(x',y')$. The unperturbed Keplerian described by
  $\vec{r}_b$ is defined with respect to the $(x',y')$ plane by the angles $\Omega$ (relative node longitude) and 
  $i$ (relative inclination). 
  The observer's frame $(x,y,z)$ with $z$ being the line of sight is obtained by rotating the auxiliary 
  frame $(x',y',z')$ around the $x$ axis by an angle $I$ (inclination of $\vec{r}$'s orbit with respect to $(x,y)$).
  Moreover, $S=\angle(\vec{r},\vec{r}_ {b})$, 
  $\theta=\angle({\hat{x}},\vec{r})$, and $\theta_{b}=\angle({\hat{x}},\vec{r}_{b})$.
  \label{fig1}}
\end{figure}

We consider the framework of the general three-body problem: we assume that
a star with mass $M_{\star}$ has a nearby binary system with masses $M_{1}$ and $M_{2}$ (Fig.~\ref{fig1}).

Following \citet{Morais&Correia2008}, we use Jacobi coordinates, which are the inter-binary distance, $\vec{r_b}$, and
the distance, $ \vec{r} $, from the star $ M_{\star} $ to the binary's center of mass (see Fig.~\ref{fig1}).
We also assume that $ \rho = |\vec{r_b}| / |\vec{r}| \ll 1$.

In the auxiliary frame (Fig.~1), $\vec{r}=(x',y',z')$, where
\begin{eqnarray}
\label{xyz}
x' &=& r \cos{\theta} \ , \nonumber \\ 
y' &=& r \sin{\theta}  \ , \nonumber \\
z' &=& 0 \ ,
\end{eqnarray}
and $\vec{r}_{b}=(x'_{b},y'_{b},z'_{b})$, where
\begin{eqnarray}
\label{xyzb}
x'_{b} &=& r_{b} (\cos{\Omega}\cos(\theta_{b}-\Omega)-\sin{\Omega}\sin(\theta_{b}-\Omega)\cos{i}) \ , \nonumber \\
y'_{b} &=& r_{b} (\sin{\Omega}\cos(\theta_{b}-\Omega)+\cos{\Omega}\sin(\theta_{b}-\Omega)\cos{i}) \ , \nonumber \\
z'_{b} &=& r_{b} \sin(\theta_{b}-\Omega) \sin{i} \ ,
\end{eqnarray}
where $\theta$ and $\theta_b$ are the orbits' true longitudes, $\Omega$ is the relative node longitude, and $i$ is the relative inclination.

The transformation from the auxiliary frame $(x',y',z')$ to the observer's frame $(x,y,z)$ is
\begin{eqnarray}
\label{transfxyz}
\hat{x} &=& \hat{x'} \ , \\ \nonumber
\hat{y} &=& \cos{I}\,\hat{y'}-\sin{I}\,\hat{z'} \ , \\ \nonumber
\hat{z} &=& \sin{I}\,\hat{y'}+\cos{I}\,\hat{z'} \ .
\end{eqnarray}

Moreover, we define the angle $S=\angle(\vec{r},\vec{r}_ {b})$ which is given by
\begin{equation}
\label{coS0}
\cos{S} = \hat{r} \cdot \hat{r_{b}} \ ,
\end{equation}
where $\hat{r}$ is the versor of $\vec{r}$ and $\hat{r_{b}}$ is the versor of $\vec{r_{b}}$.

Jacobi coordinates are useful to this problem because the unperturbed motion is a composition of two Keplerian ellipses. As seen in \citet{Morais&Correia2008}, including terms up to first order in $\rho$, we have
\begin{eqnarray}
\ddot{\vec{r}}_b=-G\frac{M_{1}+M_{2}}{r_{b}^2} \left( \hat{r}_{b}+
	      \epsilon_{b} (3 \cos{S} \hat{r}-\hat{r}_{b}) \right) \ ,
\end{eqnarray}
where, unless $M_{\star} \gg M_{1}+M_{2}$,
\begin{equation}
\epsilon_{b}= \frac{ M_{\star}}{M_{1}+M_{2}} \rho^3 \ll 1
\end{equation}
and
\begin{eqnarray}
\label{eqr}
\ddot{\vec{r}}& = & -G\frac{ (M_{1}+M_{2} +M_{\star})}{r^2} \times \nonumber \\ 
              & & \left[ \hat{r }+ \epsilon \left( (-3/2+15/2 \cos^2{S}) \hat{r} - 3 \cos{S} \hat{r}_{b} \right) \right] \ ,
\end{eqnarray}
where
\begin{equation}
\epsilon=\frac{M_{1} M_{2}}{( M_{1}+M_{2} )^2} \rho^2 \le \frac{\rho^2}{4} \ll 1 \ .
\end{equation}
Thus we can write
$\vec{r}_{b} = \vec{r}_{b0}+\epsilon_{b} \vec{r}_{b1}$ \ ,
where
\begin{equation}
\ddot{\vec{r}}_{b0}=-G\frac{M_{1}+M_{2}}{r_{b0}^2} \hat{r}_{b0} \ ,
\end{equation}
and 
$\vec{r} = \vec{r}_{0}+\epsilon \vec{r}_{1}$ \ ,
where
\begin{equation}
\ddot{\vec{r}}_{0} = -G\frac{ (M_{1}+M_{2} +M_{\star})}{r_{0}^2} \hat{r}_{0} \ .
\end{equation}
Therefore
the 0th order solution, $\vec{r}_{b0}$, is a Keplerian ellipse with
constant semi-major axis, $a_b$, and frequency 
\begin{equation}
\label{freqbin}
n_{b}=\sqrt{\frac{G ( M_{1}+M_{2} )}{a_{b}^3}} \ ,
\end{equation}  
and
the 0th order solution,  $\vec{r}_{0}$, is a Keplerian ellipse with
constant semi-major axis, $a$, and frequency 
\begin{equation}
\label{freqstar}
n=\sqrt{\frac{G (M_{1}+M_{2} + M_{\star})}{a^3}}  \ .
\end{equation}
Furthermore, from Eqs.~(\ref{freqbin}) and (\ref{freqstar}) we see that in general $n \ll n_{b}$  since $ a_b \ll a $ 
(this is true unless $M_{\star} \gg M_{1}+M_{2}$).

Now, the distance of the star to the triple system's center of mass is
\begin{equation}
\vec{r}_{\star} = \frac{M_{1}+M_{2}}{M_{1}+M_{2}+M_{\star}} \, \vec{r}  \ ,
\end{equation}
hence from Eq.~(\ref{eqr}) we have
\begin{eqnarray}
\ddot{\vec{r}}_{\star}& = & -G \frac{M_{1}+M_{2}}{r^2} \times \nonumber \\ 
              & & \left[ \hat{r }+ \epsilon \left( (-3/2+15/2 \cos^2{S}) \hat{r} - 3 \cos{S} \hat{r}_{b} \right) \right] \ .
\end{eqnarray}
Since $r^{-2}=r_{0}^{-2} (1+O(\epsilon))$, $\hat{r}=\hat{r}_{0} (1+O(\epsilon))$ and 
$\hat{r}_{b}=\hat{r}_{b0} (1+O(\epsilon_{b}))$, 
we obtain an approximation to the previous equation that is accurate up to $O(\epsilon)$, i.e.
\begin{eqnarray}
\label{eqrstar}
\ddot{\vec{r}}_{\star } &=& -G\frac{ M_{1}+M_{2}}{r^2}\hat{r} \\ \nonumber
&& -G\frac{ M_{1}+M_{2} }{r_{0}^2} \epsilon \left[ 
\left(-\frac{3}{2}+\frac{15}{2} \cos^2{S} \right) \hat{r}_{0} - 3 \cos{S} \hat{r}_{b0} \right] \ , 
\end{eqnarray}
whose solution, $\vec{r}_{\star}=\vec{r}_{\star0}+ \epsilon \vec{r}_{\star1}$, is thus
a combination of a Keplerian ellipse (with frequency $n$), $\vec{r}_{\star0}$, and a perturbation term 
\begin{equation}
\label{eqrstar1}
\ddot{\vec{r}}_{\star1 } = -G\frac{ M_{1}+M_{2} }{r_{0}^2} \left[ 
\left(-\frac{3}{2}+\frac{15}{2} \cos^2{S} \right) \hat{r}_{0} - 3 \cos{S} \hat{r}_{b0} \right] \ .
\end{equation} 

The radial velocity is the projection of the star's barycentric velocity along the line of sight 
(defined as the $z$-axis in Fig.~\ref{fig1}), i.e., $V_{r}=\dot{z}_{\star}$. 
Therefore, we only need to compute the $z$ component of the first order correction, 
$\epsilon \vec{r}_{\star 1}$ (Eq.~(\ref{eqrstar1})), i.e.
\begin{eqnarray}
\label{eqz}
\ddot{z}_{\star} & = & -G \frac{M_{1}+M_{2}}{r_{0}^2} \epsilon \times \nonumber \\
              & &  \left( \left( \frac{9}{4}+\frac{15}{4} \cos(2 S) \right)
 \hat{r}_{0} \cdot \hat{z} - 3 \cos{S} \hat{r}_{b0} \cdot \hat{z} \right) \ .
\end{eqnarray}

Additionally, from Eqs.~(\ref{xyz}),~(\ref{xyzb}),~(\ref{transfxyz}), and~(\ref{coS0}) we have
\begin{eqnarray}
\label{z0}
\hat{r}_{0} \cdot \hat{z} &=& \sin(\theta) \sin{I} \ ,\\
\label{zb0}
\hat{r}_{b0} \cdot \hat{z} &=& \sin(\theta_{b}) \sin{I}+\sin(\theta_{b}-\Omega)\times \\ \nonumber
                           && \left( \sin{i}\,\cos{I}-(1-\cos{i})\cos{\Omega}\,\sin{I} \right) \, \\
\label{coS}
\cos{S} &=& \hat{r} \cdot \hat{r}_{b} \approx \hat{r}_{0} \cdot \hat{r}_{b0}  \nonumber \\
&=& \cos(\theta_{b}-\theta)-(1-\cos{i})\,\sin(\theta-\Omega)\,\sin(\theta_{b}-\Omega) \ .
\end{eqnarray}

\subsection{Star and binary on circular orbits}

In this case , $r_{0}=a$, $r_{b0}=a_b$, $\theta=\lambda=n\,t+\lambda_0$, $\theta_b =\lambda_b = n_{b}\,t+\lambda_{b0}$,
where $\lambda$ and $\lambda_b$ are the orbits' mean longitudes.
Thus Eq.~(\ref{eqz}) becomes
\begin{eqnarray}
\label{eqz3Dcircular}
\ddot{z}_{\star} &=& \frac{3}{4}\,\delta\,n_{b}^2\,\sin{I}\times
 \left[  4 (1-3\,{ \cos{i}}^{2}) 
\sin( n\,t +\lambda_0 ) \right. \nonumber \\
&&+ 2 (1-{ \cos{i}}^{2})
 \sin( n\,t +\lambda_0 -2\,\Omega ) \nonumber \\
&&- 10 (1-{\cos{i}}^{2})
 \sin( 3\,n\,t +3\,\lambda_0 -2\,\Omega) \nonumber \\
&&+ 5 (1+ \cos{i})^{2}  
\sin( (2\, n_b -3\,n)\,t +2\,\lambda_{b0}-3\,\lambda_0 ) \nonumber \\
&&  -5 (1-{\cos{i}})^{2} ) 
\sin( (2\, n_b + 3\,n)\,t +2\,\lambda_{b0} +3\,\lambda_0 -4\,\Omega ) \nonumber \\
&& - (1+{\cos{i}} )^{2} 
\sin( (2\, n_b -n)\,t +2\,\lambda_{b0} -\lambda_0 )  \nonumber \\
&&+ 6 ( 1-{\cos{i}}^{2} )
 \sin( (2\, n_b -n)\,t +2\,\lambda_{b0} -\lambda_0 -2\,\Omega ) \nonumber \\
&& + (1- \cos{i})^{2} 
\sin( (2\, n_b +n)\,t +2\,\lambda_{b0} +\lambda_0 -4\,\Omega )  \nonumber \\
&& \left. - 6 (1-{ \cos{i}}^{2} )
 \sin( 2\, n_b +n)\,t +2\,\lambda_{b0} +\lambda_0 -2\, \Omega ) \right] \nonumber \\
&& + 6\,\delta\,n_{b}^2\,\cos{I}\times 
 \left[ 2\, \sin{i}\,\cos{i}\,
\sin ( n\,t +\lambda_0 -\Omega) \right. \nonumber \\ 
&& +  \sin{i} (1+ \cos{i})
 \sin( (2\, n_b -n)\,t +2\,\lambda_{b0} -\lambda_0 -\Omega ) \nonumber \\
&& \left.  +\sin{i} (1-\cos{i})
 \sin( (2\, n_b +n)\,t +2\,\lambda_{b0} +\lambda_0 -3\,\Omega) \right] \ ,
\end{eqnarray}
where
\begin{equation}
\label{delta}
\delta=\frac{M_{1} M_{2}}{8 ( M_{1}+M_{2} )^2} \left( \frac{a_b}{a} \right)^4 a_{b} \,
\end{equation}
as defined in \citet{Morais&Correia2008}.

The right-hand side of Eq.~(\ref{eqz3Dcircular}) is a sum of periodic terms, hence $V_{r}=\dot{z}_{\star}$ is obtained by integrating them with respect to time. Therefore, $\dot{z}_{\star}$ is a linear combination of six periodic terms with frequencies $n$, $3\,n$, $2\,n_{b}-n$, $2\,n_{b}+n$, $2\,n_{b}-3\,n$, and $2\,n_{b}+3\,n$. 

In the coplanar case ($i=0$), we recover the solution obtained in \citet{Morais&Correia2008} 
\begin{eqnarray}
\dot{z}_{\star} & = &  6 \delta \sin{I} \frac{n_{b}^2}{n} \cos(n\,t+\lambda_{0}) \nonumber \\
&&  -15 \delta \sin{I} \frac{n_{b}^2}{2\,n_{b}-3\,n}  \cos((2\,n_{b}-3\,n)\,t+2\,\lambda_{0}-3\,\lambda_{b0}) \nonumber \\
&& +3 \delta \sin{I} \frac{n_{b}^2}{2\,n_{b}-n} \cos((2\,n_{b}-n)\,t+2\,\lambda_{0}-\lambda_{b0}) \,
\end{eqnarray}
i.e., $\dot{z}_{\star}$ is a composition of three periodic terms with frequencies $n$, $2\,n_{b}-n$, and $2\,n_{b}-3\,n$.

We note that when $i=0$ (coplanar case) the angle $I$ defines the projection of the star's orbit
along the line of sight and thus $\ddot{z}$ (and $\dot{z}$) scale with $\sin{I}$.
However, when $i\neq 0$ the angle that defines the projection of the star's orbit along the line of sight depends 
on $I$, $i$, and $\Omega$. When $i\neq 0$,
the amplitudes associated with the frequencies $n$ and $2\,n_{b}\pm n$ indeed do not scale with $\sin{I}$ (cf.~last three terms in
Eq.~(\ref{eqz3Dcircular})). In particular, when $I=0$ (pole-on configuration) and if $i\neq 0$ we can still detect these terms (frequencies $n$ and $2\,n_{b}\pm n$), which are due to the binary's effect on the star.

In Fig.~(2), we show the normalized amplitudes ($A$) associated with the frequencies $2\,n_{b}\pm n$ and $2\,n_{b}\pm 3\,n$,
as a function of $i$ for $\Omega=0$, $I=90^\circ$ (Fig.~2, top) and $\Omega=90^\circ$, $I=90^\circ$ (Fig.~2, bottom).
In Fig.~(3), we show the equivalent picture for $I=0$. The true amplitudes are $A\,\delta\,n_{b}$.

At $I=90^\circ$ (equator-on configuration) and small $i$, the term of frequency $2\,n_{b}-3\,n$ is dominant but as $i$ increases its amplitude decreases, while the amplitudes of the terms with frequencies $2\,n_{b}+3\,n$ and $2\,n_{b}\pm n$ increase until they are all approximately equal at $i=90^\circ$.
At $I=0$ (pole-on configuration), only the terms with frequencies $n$ and $2\,n_{b}\pm n$ appear (cf.~Eq.~(\ref{eqz3Dcircular})), hence we cannot detect the star's motion around the binary's center of mass but if $i\neq 0$ we can still detect the binary's effect on the star. 
Moreover, the amplitudes associated with the frequencies $2\,n_{b}\pm 3\,n$ are independent of $\Omega$ at any value of $I$ while the amplitudes associated with the frequencies $2\,n_{b} \pm n$ depend  on the orbits' intersection $\Omega$ except when $I=0$ (cf.~Eq.~(\ref{eqz3Dcircular})).       
The pictures for $90^\circ > i \ge 0^\circ$ (star and binary on prograde orbits) and $180^\circ \ge i >90^\circ$ (binary on a retrograde orbit) are symmetrical except that the frequency $n$ is replaced with $-n$, which corresponds to inverting the star's (or binary's) motion. 

\begin{figure}
   \centering
    \includegraphics[width=6cm,angle=0]{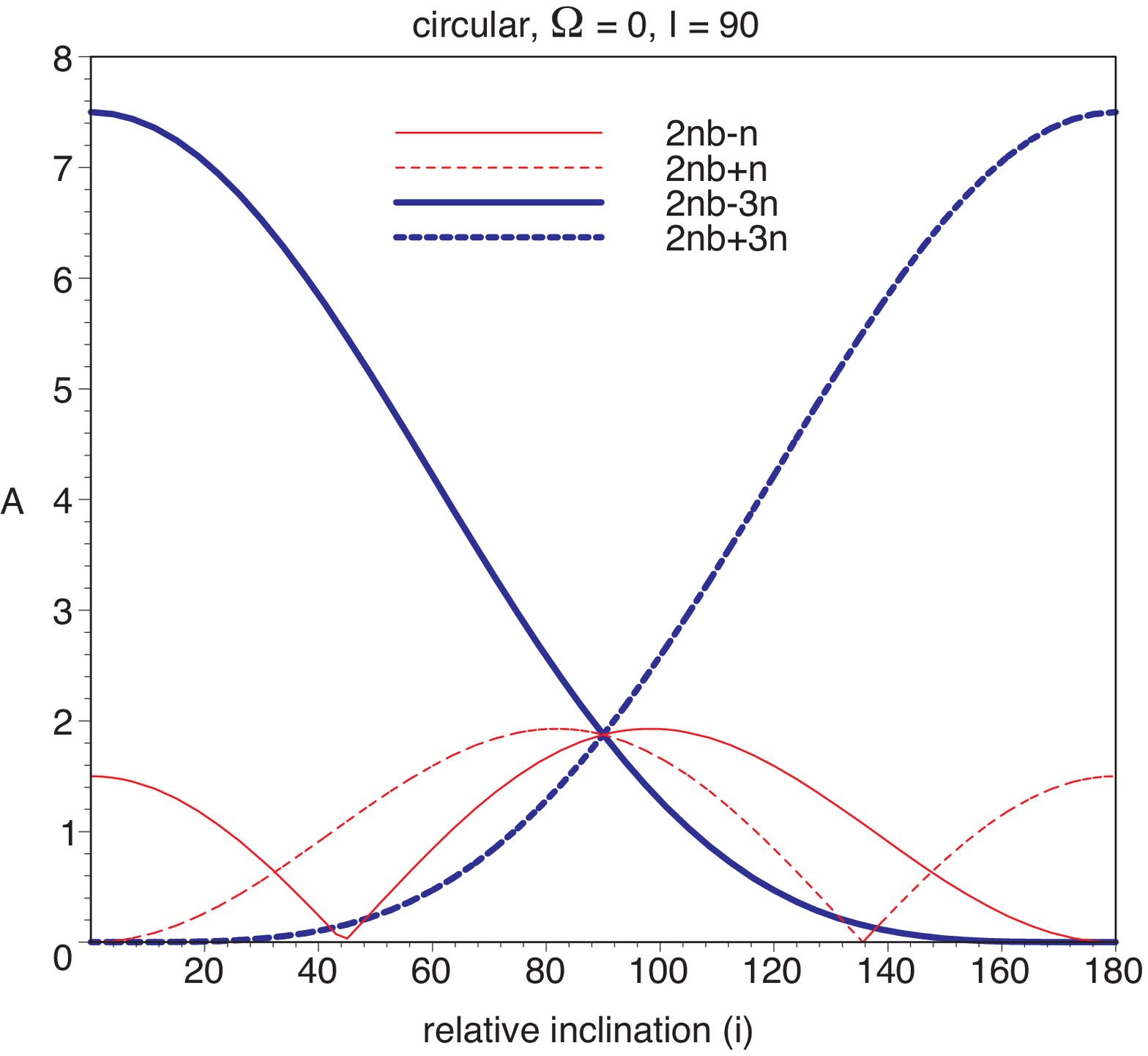}
    \includegraphics[width=6cm,angle=0]{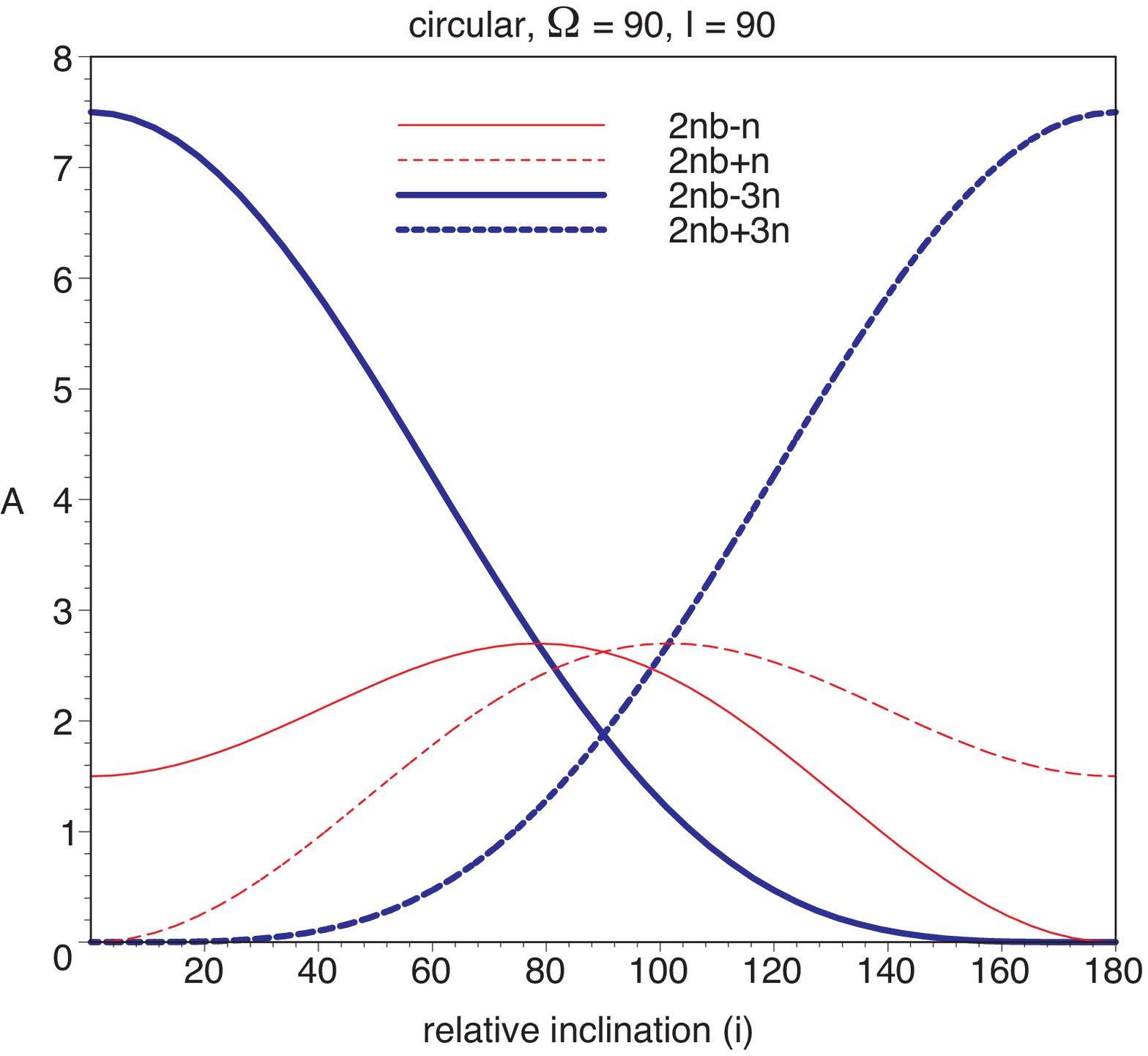}
  \caption{Circular case with $I=90^\circ$ (equator-on configuration). Normalized amplitudes $A$ of periodic terms 
as function of the relative inclination $i$. 
These are obtained by integrating Eq.~(\ref{eqz3Dcircular}) with respect to time.
The true amplitudes are $A\,\delta\,n_{b}$.}
\end{figure} 

\begin{figure}
   \centering
    \includegraphics[width=6cm,angle=0]{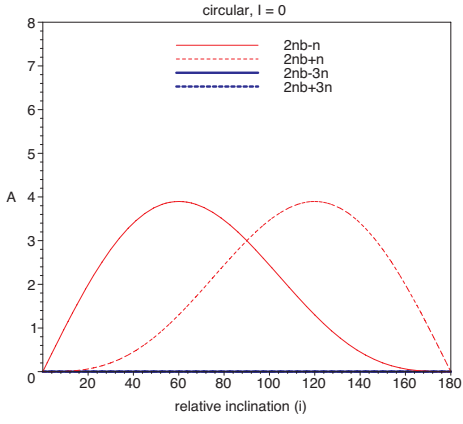}
  \caption{Circular case with $I=0$ (pole-on configuration). Normalized amplitudes $A$ of periodic terms 
as function of the relative inclination $i$. These are obtained by integrating Eq.~(\ref{eqz3Dcircular}) with respect to time. The true amplitudes are $A\,\delta\,n_{b}$.}
\end{figure}

\subsection{Star and binary on eccentric orbits}

We assume that the Keplerian orbit, $\vec{r}_{0}$, has a semi-major axis $a$, eccentricity $e$, and true longitude $\theta=f+\varpi$ (where $f$ is the true anomaly and $\varpi$ is the longitude of periastron) and that the Keplerian orbit, $\vec{r}_{b0}$, has semi-major axis $a_b$, eccentricity $e_b$, and true longitude $\theta_{b}=f_{b}+\varpi_{b}$ (where $f_b$ is the true anomaly and $\varpi_b$ is the longitude of periastron). Thus Eq.~(\ref{eqz}) becomes
\begin{eqnarray}
\label{eqzeccentric}
\ddot{z}_{\star} & = & -G \frac{M_{1}+M_{2}}{a^2} \epsilon_{0} \left( \frac{a}{r_0} \right)^4 \left( \frac{r_{b0}}{a_b} \right)^2 \times \nonumber \\
              & & 
\left( \left( \frac{9}{4}+\frac{15}{4} \cos(2 S) \right)
 \hat{r}_{0} \cdot \hat{z} - 3 \cos{S} \hat{r}_{b0} \cdot \hat{z} \right) \ ,
\end{eqnarray}
where
\begin{equation}
\epsilon_{0}=\frac{M_{1} M_{2}}{( M_{1}+M_{2} )^2} \left( \frac{a_b}{a} \right)^2 \ .
\end{equation}
 
We can express the orbital solutions, $\vec{r}_{0}$ and $\vec{r}_{b0}$, as Fourier series of the mean anomalies
$M=n\,t+M_{0}$ and $M_{b}=n_{b}\,t+M_{b0}$, respectively, also known as elliptic expansions.
In particular, we obtain expressions for $(a/r_{0})^4$ and $(r_{b0}/a_{b})^2$ using  elliptic expansions for 
$r/a$ and $r_{b}/a_{b}$ (e.g.~\citet{murray_dermott1999}). 
We obtain similar expressions  for $\hat{r}_{0} \cdot \hat{z}$ (Eq.~(\ref{z0})), $\hat{r}_{b0} \cdot \hat{z}$ (Eq.~(\ref{zb0})), and $\cos{S}$ (Eq.~(\ref{coS}))  using elliptic expansions for  $\cos(f)$, $\sin(f)$, $\cos(f_b)$,
and $\sin(f_b)$  (e.g.~\citet{murray_dermott1999}).

\subsubsection{Coplanar motion}

In this case, $i=0$, thus from Eqs.~(\ref{z0}),~(\ref{zb0}), and~(\ref{coS}) 
\begin{eqnarray}
\hat{r}_{0} \cdot \hat{z} &=& \sin(\theta) \sin{I} \ , \\
\hat{r}_{b} \cdot \hat{z} &=& \sin(\theta_{b}) \sin{I} \ , \\
\cos{S} &=& \cos(\theta_{b}-\theta) \ ,
\end{eqnarray}
where $\theta=f+\varpi$ and $\theta_{b}=f_{b}+\varpi_{b}$. 

The next step is to replace these expressions into  Eq.~(\ref{eqzeccentric}) using elliptic expansions for
$(a/r_{0})^4$, $(r_{b0}/a_{b})^2$, $\sin(\theta)$, $\sin(\theta_{b})$, $\cos(\theta_{b}-\theta)$,
and $\cos(2\,(\theta_{b}-\theta))$. We start by using elliptic expansions up to first order in the eccentricities.

As in \citet{Morais&Correia2008}, our aim is to obtain a solution to Eq.~(\ref{eqzeccentric}) that is a 
linear combination of periodic terms. However, when $e \ne 0$ we have a constant term of the type 
$e \sin(\varpi)$ appearing in
$\ddot{z}_{\star}$. We can remove this secular term by rewriting Eq.~(\ref{eqrstar1}) as
\begin{equation}
\label{eqrstar1new}
\ddot{\vec{r}}_{\star 1} =  -G \frac{M_{1}+M_{2}}{r_{0}^2} \vec{p}+\ddot{\vec{r'}}_{\star 1} \ ,
\end{equation} 
where the vector $\vec{p}=(p_{x},p_{y},p_{z})$ has radius $(3/4)\,e$, and is aligned with the periapse of the  unperturbed orbit, $\vec{r}_0$, i.e.
\begin{eqnarray}
p_{x} &=& (3/4) e \cos(\varpi) \ , \\
p_{y} &=& (3/4) e \sin(\varpi)\cos{I} \ , \\
p_{z} &=& (3/4) e \sin(\varpi) \sin{I} \ ,
\end{eqnarray}
and
\begin{eqnarray}
\label{eqrstar1reduced}
\ddot{\vec{r'}}_{\star 1} &=& G \frac{M_{1}+M_{2}}{r_{0}^2}  \vec{p}
-G \frac{M_{1}+M_{2}}{r_{0}^2}  \times \nonumber \\
              & &  \left( \left( \frac{9}{4}+\frac{15}{4} \cos(2 S) \right)
 \hat{r}_{0} - 3 \cos{S} \hat{r}_{b0} \right) \ .
\end{eqnarray} 
Now, the first term in Eq.~(\ref{eqrstar1new}) can be interpreted as the slow rotation
(frequency $\sim\epsilon_{0}^{1/2}\,n \ll n \ll n_b$) of the unperturbed orbit $\vec{r}_0$'s periapse described by the vector $\vec{p}$. 
Therefore, we can concentrate in obtaining the solution to Eq.~(\ref{eqrstar1reduced}).

The $z$ component of the first order correction, 
$\epsilon \vec{r'}_{\star 1}$ given by Eq.~(\ref{eqrstar1reduced}), becomes
\begin{eqnarray}
\label{eqz2Deccentric1}
\ddot{z}_{\star} &=&  3\,\delta\,n_{b}^{2}\sin{I} \times \left[ -2\,\sin \left( n\,t+\lambda_{0} \right) \right. \nonumber \\
&& -\sin \left( (2\,n_{b}-n)\,t +2\,\lambda_{b0}-\lambda_0 \right) \nonumber \\ 
&& +5\,\sin \left( (2\,n_{b}-3\,n)\,t +2\,\lambda_{b0}-3\,\lambda_{0} \right) \nonumber \\
&& -15\,e_{b}\,\sin \left((n_{b}-3\,n)\,t+\lambda_{b0}-3\,\lambda_{0}+\varpi_{b} \right) \nonumber \\ 
&& -e_{b}\,\sin \left((3\,n_{b}-n)\,t+3\,\lambda_{b0}-\lambda_{0} -\varpi_{b} \right) \nonumber \\ 
&& +5\,e_{b}\,\sin \left((3\,n_{b}-3\,n)\,t+3\,\lambda_{b0}-3\,\lambda_{0} -\varpi_{b} \right) \nonumber \\ 
&& +2\,e_{b}\,\sin \left((n_{b}+n)\,t+\lambda_{b0}+\lambda_{0} -\varpi_{b} \right) \nonumber \\
&& +3\,e_{b}\,\sin \left((n_{b}-n)\,t +\lambda_{b0}-\lambda_{0} +\varpi_{b} \right) \nonumber \\
&& -2\,e_{b}\,\sin \left((n_{b}-n)\,t+\lambda_{b0}-\lambda_{0}-\varpi_{b} \right) \nonumber \\ 
&& -6\,e\sin \left(2\,n\,t+2\,\lambda_{0} -\varpi \right) \nonumber \\
&& -e\sin \left(2\,n_{b}\,t+2\,\lambda_{b0}-\varpi \right) \nonumber \\
&& +25\,e\sin \left((2\,n_{b}-4\,n)\,t+2\,\lambda_{b0}-4\,\lambda_{0}+\varpi \right) \nonumber \\
&& -3\,e\sin \left((2\,n_{b}-2\,n)\,t+2\,\lambda_{b0}-2\,\lambda_{0}+\varpi \right) \nonumber \\
&& \left. -5\,e\sin \left((2\,n_{b}-2\,n)\,t+2\,\lambda_{b0}-2\,\lambda_{0}-\varpi \right) \right] \ ,
\end{eqnarray}
where $\lambda_{0}=M_{0}+\varpi$ and $\lambda_{b0}=M_{b0}+\varpi_{b}$.

The right-hand side of Eq.~(\ref{eqz2Deccentric1}) is a sum of periodic terms, hence $V_{r}=\dot{z}_{\star}$ is obtained by integrating these with respect to time.
Therefore, when including terms up to first order in the eccentricities, $\dot{z}_{\star}$ is a composition of 12 periodic terms with frequencies: $n$, $2\,n_{b}-n$ and $2\,n_{b}-3\,n$ (zero order); $n_{b}-3\,n$, $3\,n_{b}-n$, $3\,n_{b}-3\,n$, $n_{b}+n$ and  $n_{b}-n$  (first order: $e_b$); $2\,n$, $2\,n_{b}$, $2\,n_{b}-4\,n$ and $2\,n_{b}-2\,n$ (first order: $e$). 
The amplitudes of the terms with frequencies $n_{b}-n$ and $2\,n_{b}-2\,n$ depend on $\varpi_b$ and $\varpi$, respectively. 

We can easily extend our theory by including higher-order eccentricity terms in the elliptic expansions\footnote{ We performed these expansions using the computer algebra software Maple.} (i.e., Fourier series of the Keplerian orbital solutions $\vec{r}_{0}$ and $\vec{r}_{b0}$). In particular, if we include terms up to second order in the eccentricities, there are additional periodic terms with frequencies:  $3\,n$, $2\,n_{b}+n$, $4\,n_{b}-n$, $4\,n_{b}-3\,n$ (second order: $e_{b}^2$); $3\,n$, $2\,n_{b}+n$, $2\,n_{b}-5\,n$ (second order: $e^2$);
$3\,n_{b}-2\,n$, $n_b$, $3\,n_{b}$, $n_{b}-4\,n$, $n_{b}+2\,n$, $n_{b}-2\,n$, $3\,n_{b}-4\,n$ (second order:
$e\,e_b$). Moreover, the amplitudes associated with the frequencies $2\,n_{b}-n$ and $2\,n_{b}-3\,n$ (zero order) have corrections of second order in the eccentricities. 

In Fig.~(4), we show the largest normalized amplitudes ($A>2$) of the periodic terms that appear up to fourth order in the eccentricities (except harmonics of $n$) as functions of $e_b$ when $e=0.1$. The expansion in $e_b$ converges rapidly thus it is not necessary to include higher order terms at least up to $e_{b}=0.4$.  When $e_b$ is small, the term with frequency $2\,n_{b}-3\,n$ is dominant but when $e_{b}>0.16$, the term with frequency $n_{b}-3\,n$ has the largest amplitude. Other important terms have frequencies $2\,n_{b}-4\,n$ and $n_{b}-4\,n$.  

In Fig.~(5), we show the largest normalized amplitudes ($A>2$) of the periodic terms that appear up to 10th order in the eccentricities (except harmonics of $n$) as functions of $e$ when $e_{b}=0$. We include corrections in the amplitudes up to 12th order in the eccentricities. These high order expansions are necessary to obtain results valid up to $e=0.4$ since the expansion in $e$ converges slowly.
When $e$ is small, the term with frequency $2\,n_{b}-3\,n$  has the largest amplitude but as $e$ increases, the dominant term becomes $2\,n_{b}-4\,n$ ($e>0.18$), $2\,n_{b}-5\,n$ ($e>0.28$), and $2\,n_{b}-6\,n$ ($e>0.36$). As $e$ increases, the number of terms with frequencies near $2\,n_{b}$ and similar amplitudes increases.

\begin{figure}
   \centering
    \includegraphics[width=8cm,angle=0]{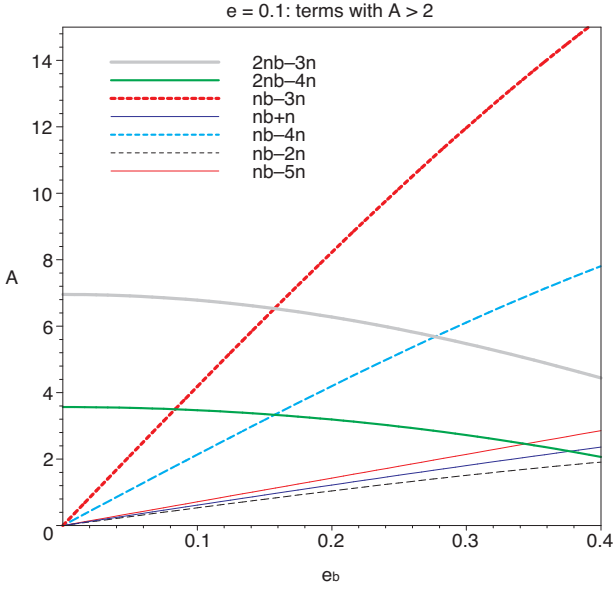}
  \caption{Coplanar case ($i=0$) with star on an eccentric orbit ($e=0.1$).
Normalized amplitudes $A$ of periodic terms, as function of the binary's eccentricity $e_b$.
We show all periodic terms that appear up to fourth order in the eccentricities and reach at least $A=2$.
The true amplitudes are $A\,\delta\,n_{b}\,\sin{I}$.}
\end{figure}   

\begin{figure}
   \centering
    \includegraphics[width=8cm,angle=0]{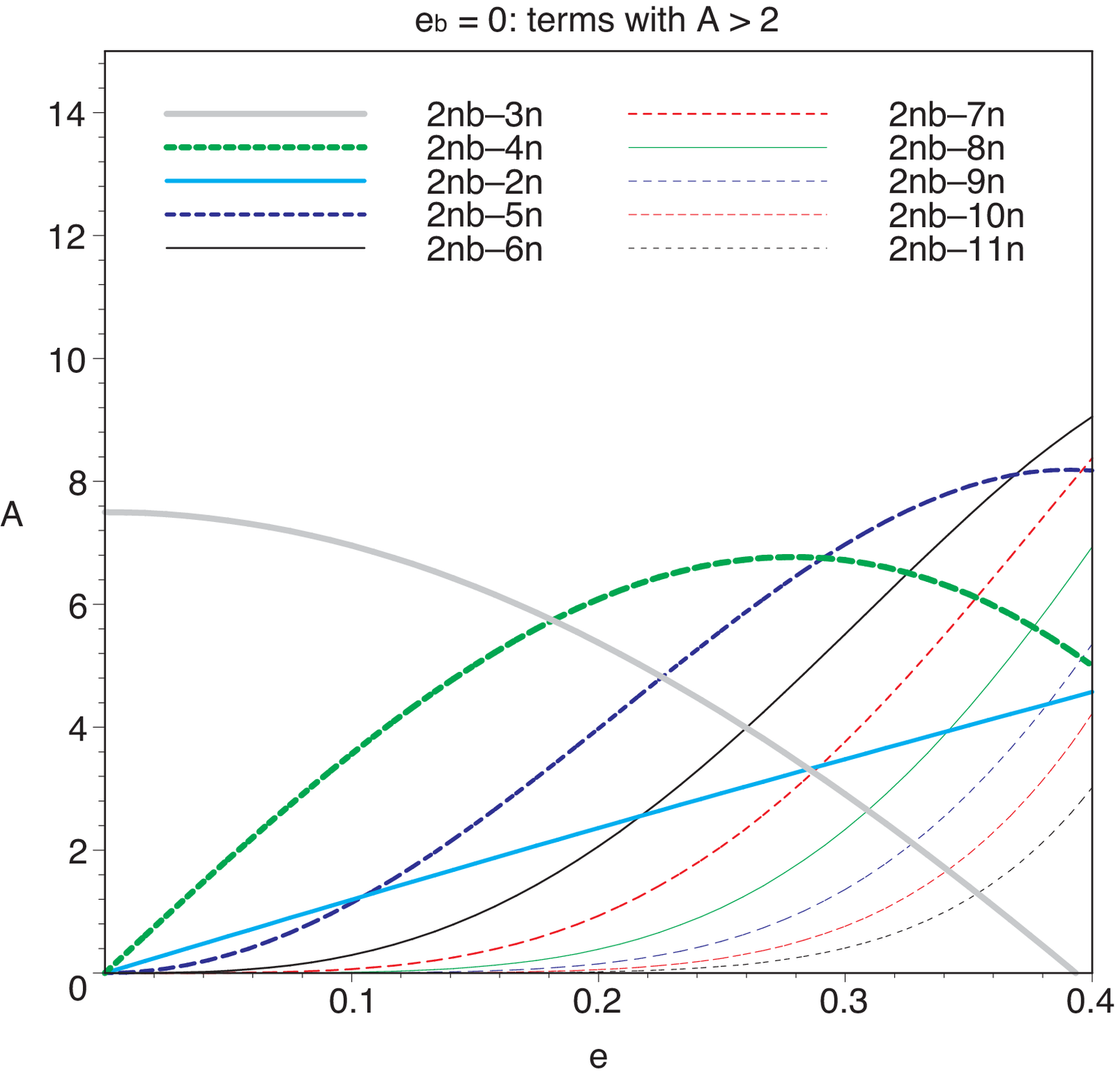}
  \caption{Coplanar case ($i=0$) with star on an eccentric orbit ($e=0.1$).
Normalized amplitudes $A$  of periodic terms, as function of the star's eccentricity $e$.
We show all periodic terms that appear up to 10th order in the eccentricities and reach at least $A=2$.
The true amplitudes are $A\,\delta\,n_{b}\,\sin{I}$.}
\end{figure}   

\subsubsection{Non-coplanar motion}

When $i\neq 0$, we must follow the same procedure described in the previous section, expressing the orbital solutions $\vec{r}_{0}$ and $\vec{r}_{b0}$  using elliptic expansions (e.g.~\citet{murray_dermott1999}), but now using the general expressions for $\hat{r}_0 \cdot \hat{z}$, $\hat{r}_{b0} \cdot \hat{z}$, and $\cos{S}$ (Eqs.~(\ref{z0}),~(\ref{zb0}), and~(\ref{coS})). 

We can show that, including terms up to first order in the eccentricities, $\ddot{z}_{\star}$ (and $\dot{z}_{\star}$) are linear combinations of 21 periodic terms with frequencies $n$, $2\,n$, $3\,n$, $4\,n$, $2\,n_{b}$, $2\,n_{b}\pm n$, $2\,n_{b}\pm 3\,n$, $n_{b}\pm n$, $n_{b} \pm 3\,n$, $3\,n_{b} \pm n$, $3\,n_{b} \pm 3\,n$, $2\,n_{b}\pm 4\,n$, and $2\,n_{b}\pm 2\,n$. We note that, except for the additional harmonics of $n$, these include the set of frequencies already present in the coplanar eccentric case plus an additional set of frequencies obtained from the previous set by replacing $n$ with $-n$, which as noted previously corresponds to inverting the star's (or binary's) motion.

We already described how the terms with frequencies  
$2\,n_{b} \pm n$, and $2\,n_{b} \pm 3\,n$ (circular case) behaved with $I$, $i$, and $\Omega$ (Figs.~(2),(3) and Sect.~2.2).
To understand what happens when $e$ and $e_b$ are small but non-zero, we now describe the behavior of the terms that appear to first order in the eccentricities. 

When $e_{b} \ne 0$, the frequencies $n_{b}\pm 3\,n$, $3\,n_{b} \pm n$, $n_{b} \pm n$, and $3\,n_{b} \pm 3\,n$  appear.
In Fig.~(6), we show the normalized amplitudes ($A$) as functions of $i$ for $I=90^\circ$ (equator-on configuration),
 $e_{b}=0.1$, and $\Omega=90^\circ$. 
At $I=90^\circ$ and small $i$, the term with frequency $n_{b}-3\,n$ has maximum amplitude. As the relative inclination $i$ increases, this amplitude decreases and at $i=90^\circ$ the terms  with frequencies 
$n_{b} \pm n$ have maximum amplitude.

\begin{figure}
   \centering
   \includegraphics[width=6cm,angle=0]{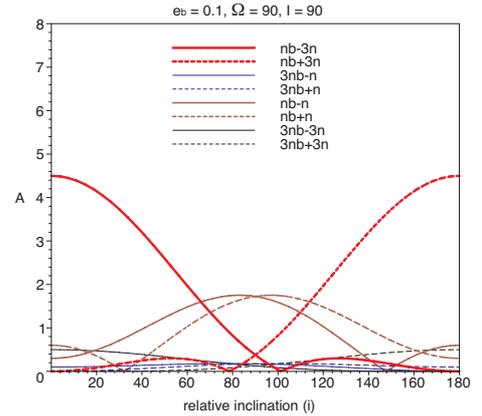}
  \caption{Eccentric case ($e_{b}=0.1$) with $I=90^\circ$. Normalized amplitudes $A$ of periodic terms 
as a function of the relative inclination $i$. The true amplitudes are $A\,\delta\,n_{b}$.}
\end{figure}

When $e \ne 0$, the frequencies $2\,n_{b}$, $2\,n_{b} \pm 2\,n$, and $2\,n_{b} \pm 4\,n$ appear. In Fig.~(7), we show the normalized amplitudes ($A$) as functions of $i$ for $I=90^\circ$ (equator-on configuration), $e=0.1$, and
$\Omega=90^\circ$.
At $I=90^\circ$ and small $i$, the term with frequency $2\,n_{b}-4\,n$ has maximum amplitude. As the relative inclination 
$i$ increases, this amplitude decreases and at $i=90^\circ$ it matches the amplitudes of the terms with frequency 
$2\,n_{b}\pm 2\,n$.

\begin{figure}
   \centering
    \includegraphics[width=6cm,angle=0]{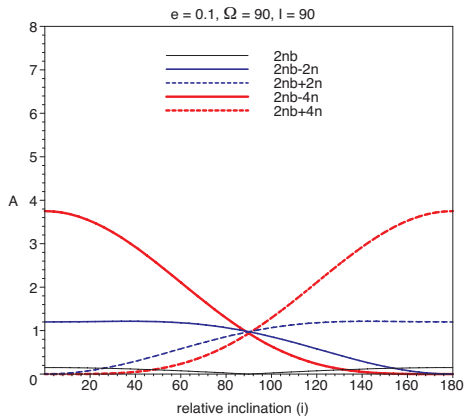}
  \caption{Eccentric case ($e=0.1$) with $I=90^\circ$. Normalized amplitudes $A$ of periodic terms 
as a function of the relative inclination $i$. The true amplitudes are $A\,\delta\,n_{b}$.}
\end{figure}  

In Figs.~(8) and~(9), we show the normalized amplitude ($A$) when $I=0$, $e_{b}=0.1$, and $e=0.1$, respectively. In this configuration ($I=0$), only the frequencies $3\,n_{b} \pm n$, $n_{b} \pm n$, $2\,n_{b}$, and $2\,n_{b} \pm 2\,n$ have non-zero amplitude.
As noted previously, in this case (pole-on configuration) we cannot detect the star's motion around the binary's center of mass but if $i\neq 0$, we can still detect the binary's effect on the star. 

\begin{figure}
   \centering
    \includegraphics[width=6cm,angle=0]{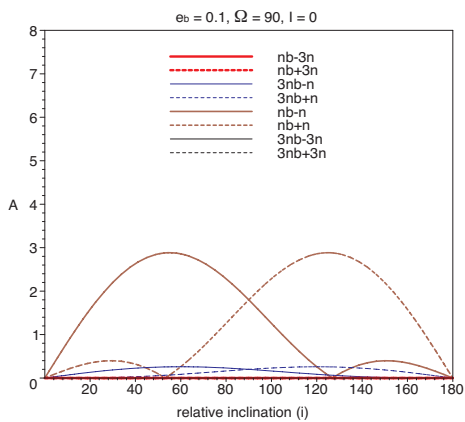}
  \caption{Eccentric case ($e_{b}=0.1$) with $I=0$. Normalized amplitudes $A$ of periodic terms 
as a function of the relative inclination $i$. The true amplitudes are $A\,\delta\,n_{b}$.}
\end{figure}  

\begin{figure}
   \centering
    \includegraphics[width=6cm,angle=0]{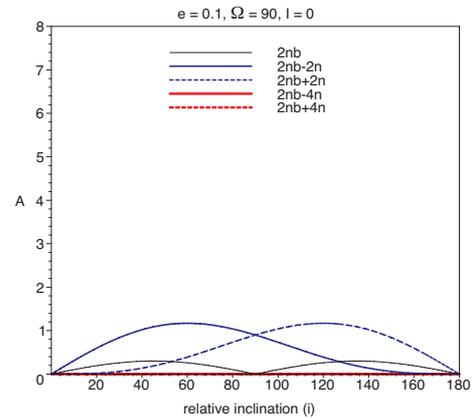}
  \caption{Eccentric case ($e=0.1$) with $I=0$. Normalized amplitudes $A$ of periodic terms 
as a function of the relative inclination $i$. The true amplitudes are $A\,\delta\,n_{b}$.}
\end{figure}  

Moreover, when $i \ne 0$ only the amplitudes associated with the frequencies $n_{b} \pm 3\,n$, 
$3\,n_{b} \pm n$, $n_{b} \pm n$, $2\,n_{b}$, and $2\,n_{b} \pm 2\,n$  depend on the relative node longitude, 
$\Omega$, since the other frequencies originate from the angles $2\,\theta_{b}\pm 3\,\theta$ and $3\,\theta$ whose amplitudes are independent of $\Omega$ (cf.~third, fourth and fifth terms in Eq.~(\ref{eqz3Dcircular})). 
We can also show that the amplitudes associated with the frequencies $n_{b} \pm 3\,n$ and $n_{b} \pm n$ depend on the binary's longitude of pericenter $\varpi_b$, while the amplitudes associated with the frequencies $2\,n_{b}$ and
$2\,n_{b} \pm 2\,n$ depend on the star's longitude of pericenter $\varpi$. 

The pictures for $90^\circ > i \ge 0^\circ$ (star and binary on prograde orbits) and $180^\circ \ge i >90^\circ$ (binary on a retrograde orbit) are symmetrical except that the frequency $n$ is replaced with $-n$, which corresponds to inverting the star's (or binary's) motion.

\section{Can a binary system mimic a planet?}

As in \citet{Morais&Correia2008}, we wish to identify the circumstances in which we can we mistake the binary's effect for that of a planet, when we do not know whether a binary is present (one or even both components may be unresolved). We have seen that the star's radial velocity will have a slow Keplerian term, which can be associated with a mass, $M_{1}+M_{2}$, located at the binary's center of mass. Hence, from the radial velocity we can infer the presence of a nearby "star", $M_{1}+M_{2}$, except when $I=0$ (pole-on configuration) where we can nevertheless detect the wobble due to the binary if $i\neq 0$. 
Moreover, the radial velocity of the star, $M_{\star}$, that has a planet companion, $M_p$, in a circular orbit with frequency $n_p$, semi-major axis $a_p$ and inclination $I_p$, and another nearby "star", $M_{1}+M_{2}$, is
\begin{equation}
\label{radialvelstarpl}
V_{r}=V_{0}+\frac{M_{p} \sin{I_p}}{M_{\star}+M_{1}+M_{2}+M_{p}} n_{p} a_{p} \cos(n_{p} t+\varpi_{p}) \ ,
\end{equation}
where $V_{0}$ includes the Keplerian motion around the "star" $M_{1}+M_{2}$, and we can neglect $M_{p}$ in the denominator of Eq.~(\ref{radialvelstarpl}) since the planet's mass is negligible with respect to the stars.

We saw above that the radial velocity of a star perturbed by a binary system is a composition of several periodic signals.
The terms that include harmonics of the frequency $n$ can in principle\footnote{This is not always true since the binary's contribution may prevent a Keplerian fit, in particular if the data set covers the long-period $T=2\pi/n$.} be associated with the star's slow Keplerian motion around the binary's center of mass (i.e., $V_0$). Therefore, we have
\begin{equation}
\label{radialvelbinary}
V_{r}=V_{0}+ \sum K_{p} \cos(n_{p} t+\varpi_{p}) \ ,
\end{equation}
where
\begin{equation}
\label{amplitudek}
K_{p} \approx A \delta n_{b}= A \frac{\sqrt{G (M_{1}+M_{2})}}{8} \frac{M_{1} M_{2}}{( M_{1}+M_{2} )^2} \left( \frac{a_b}{a} \right)^4 a_{b}^{-1/2} \ ,
\end{equation}
and the factor $A$ is a normalized amplitude that, as explained previously, can be obtained for each frequency $n_p$.
 
Each of the terms with amplitude $K_{p}$ could be identified as a planet on a circular orbit around the star with frequency 
$n_{p}=p\,n_{b} \pm q\,n$, where $p>0$ and $q \ge 0$ are integers.
However, since we have a limit on the observations' precision, we can sometimes detect only a few periodic terms with the largest amplitudes. On the other hand, since $n_{b} \gg n$, terms of equal 
$p$ have very close frequencies. Hence, if we can detect these latter terms, we associate them with planets on very close orbits that can be discarded as unstable configurations.
However, since we have a limit on the observations' resolution, we are sometimes unable to distinguish terms with very close frequencies. 

The radial velocity method of extrasolar planet's detection consists of measuring tiny Doppler shifts in the star's spectral lines caused by the star's motion. These Doppler shifts are collected over an observation timespan, $t_{obs}$, in order to obtain the star's radial velocity curve. The instrument's precision is the smallest Doppler shift that can be detected, hence the smallest detectable amplitude in the radial velocity's Fourier transform. On the other hand, the data's resolution is the smallest frequency difference that can be detected, i.e., $\delta f = 1/t_{obs}$.
Therefore, to distinguish two terms with frequencies that differ by $\Delta\,q \cdot n$, we must have 
\begin{equation}
\label{tobs}
t_{obs}>\frac{T}{\Delta\,q} \ ,
\end{equation}
where $T=2\pi/n$.

Now, if all detected frequencies are well separated (i.e., have different $p$) we may then be led to believe that the star has one or more planet companions. In that case, we apply Kepler's third law to derive the fake planets' semi-major axis, $a_p$, from the signal's frequencies, $n_p$, i.e.
\begin{equation}
\label{semajoraxis}
a_{p}=\frac{( G M_{\star})^{1/3}}{(n_{p})^{2/3}} \ ,
\end{equation}
and from Eqs.~(\ref{radialvelstarpl}) and~(\ref{radialvelbinary}) we obtain the associated minimum masses, $M_{p} \sin{I_p}$, from the signal's amplitudes $K_p$ (Eq.~(\ref{amplitudek})), i.e.
\begin{equation}
\label{minmasspl}
\frac{M_{p} \sin{I_p}}{M_{\star}+M_{1}+M_{2}}=\frac{K_{p}}{a_{p} n_{p}} \ .
\end{equation}
Moreover, from Eq.~(\ref{amplitudek}) we see that the binary's effect is more pronounced (i.e $K_p$ is larger) for 
a large ratio $a_{b}/a$, small inter-binary distance $a_b$, and a massive binary of equal stars ($M_{1}=M_{2}$).  

\subsection{Circular coplanar orbits}

This was studied in \citet{Morais&Correia2008}. In the case of coplanar circular prograde orbits, we recall that the periodic terms due to the binary's effect have frequencies $n_{p}=2\,n_{b}-n$ and $n_{p}=2\,n_{b}-3\,n$, and associated amplitudes $K_{p} \approx (3/2) n_{b} \delta \sin{I}$ and $K_{p} \approx (15/2) n_{b} \delta \sin{I}$, respectively.
 As noted previously, these signals may at first sight be identified as two planets orbiting the star with very close frequencies. However, if we could resolve these close frequencies (i.e., if $t_{obs}>T/2$, cf.~Eq.~(\ref{tobs})), we should be able to discard the planet's existence since these systems are most certainly unstable. There are nevertheless situations where we can mistake the effect  of a binary system for a planet: 
\begin{itemize}
\item We cannot resolve the two close frequencies (i.e., if $t_{obs}<T/2$, cf.~Eq.~(\ref{tobs})), hence we detect only
one signal at $\sim 2\,n_b$.
\item Owing to limited instrument's precision, we detect the signal at $2\,n_{b}-3\,n$ but not the one at $2\,n_{b}-n$ (these have amplitudes in the ratio $5/1$.)
\end{itemize}
In the latter scenario,
from Eq.~(\ref{semajoraxis}), we obtain the associated fake planet's semi-major axis
\begin{equation}
a_{p}=\frac{( G M_{\star})^{1/3}}{(2\,n_{b}-3\,n)^{2/3}}
\approx \left( \frac{M_{\star}}{4 (M_{1}+M_{2})} \right)^{1/3} a_{b} \ ,
\end{equation}
and from Eqs.~(\ref{amplitudek}) and (\ref{minmasspl}), since the motion is coplanar ($I_{p}=I$), 
we obtain the fake planet's mass
\begin{equation}
\frac{M_p}{M_{\star}+M_{1}+M_{2}}=\frac{K_{p}}{a_{p} (2\,n_{b}-3\,n)}
\approx \frac{15}{32} \frac{M_{1} M_{2}}{(M_{1}+M_{2})^2} \left( \frac{a_b}{a}\right)^4 \frac{a_b}{a_p} \ .
\end{equation}
In the former scenario, the signal's amplitude can vary by $20\%$ depending on the unresolved signals' phases. Nevertheless, these formulae are still approximately valid.

\subsection{Non-circular, non-coplanar orbits}

The case of inclined and eccentric orbits is similar. We saw that, when including terms up to first order in the eccentricities, the star's radial velocity is a composition of 21 periodic terms. Four terms are harmonics of the frequency $n$,  hence can in principle be discarded as associated with the slow Keplerian motion around the binary's center of mass. Four terms have frequencies close to $n_b$, nine terms have frequencies close to $2\,n_b$, and four terms have frequencies close to $3\,n_{b}$.  

Comparing Figs.~2,~3,~6,~7,~8, and~9 we see that the maximum radial velocity variations associated with dominant periodic terms occur when $I=90^\circ$ (equator-on configuration) and the orbits are nearly coplanar ($i \approx 0$ or $i \approx 180^\circ$). 
In the case of coplanar prograde motion ($i=0$), the dominant term has frequency $2\,n_{b}-3\,n$ (for small $e$ and $e_b$; cf.~Figs.~4 and~5) and frequency $n_{b}-3\,n$ (for large $e_b$; cf.~Fig.~4). In particular, when the binary's orbit is eccentric it is possible to detect signals at $2\,n_{b}-3\,n$ and $n_{b}-3\,n$, which can mimic two planets with orbital periods in the ratio 2/1. However, as $e$ increases the number of terms with frequencies close to $2\,n_{b}$ and similar amplitudes increases (Fig.~5), hence it should be easier to detect signals with very close frequencies.

\section{Examples}

\subsection{Theory versus simulations}

To test our model, we performed some numerical simulations of a triple system composed of main sequence stars with different spectral types G, K, and M, and masses 
$M_{\star}=M_{\odot}$, $M_{1}=0.70\,M_{\odot}$, and $M_{2}=0.35\,M_{\odot}$, respectively. 
The smaller K and M stars form a binary system with semi-major axis $a_{b}=1.1$ AU, eccentricity $e_b$, and inclination $I_b$. The G star is in a wide orbit  with semi-major axis $a=10$ AU, eccentricity $e$, and inclination $I$ around the binary's center of mass. The M star is much fainter than the K or G stars, hence it represents the unresolved component of the binary. 

We numerically integrated this triple system and computed the radial velocity of the G star (the brightest body in the system) assuming an equator-on configuration ($I=90^\circ$). From this, we simulated 100 observational data points for a time span of 4000 days (about 11 years). 
We assumed that the data were obtained with a precision limit of $0.535$ m/s, which corresponds to $A=4.46$ (cf.~Eq.~\ref{amplitudek}) and is about the highest precision that can presently be obtained (HARPS,~\citet{Harps2003}). 
We then applied the traditional techniques used to search for  planets. 

Since $t_{obs}=4000$ days and the output step is $\Delta\,t \approx 40$ days, the largest detectable frequency 
is $f_{c}=2/(\Delta\,t)=1/20$ days$^{-1}$ and the frequency resolution 
is $\Delta\,f=1/4000$ days$^{-1}$. 
We present the theoretical predictions versus simulations in Table 1.

In a first step, we can only detect the large-amplitude radial velocity variations due to the star's slow motion (with period $T=2\pi/n$) around the binary's center of mass. Once we subtract these long-term variations we are able to detect the signal due to the binary's motion, which could be mistaken for one or more planet companions to the star.

Example~1 is a circular coplanar system ($i=0$) already presented in \citet{Morais&Correia2008}, where the binary mimics a planet of approximately  20 Earth masses. Examples 2 and 3 are also circular but inclined systems with $i=30^{\circ}$ and $i=60^{\circ}$, respectively. These demonstrate that increasing the relative inclination $i$ leads to a decrease in the planet's mass $M_p$, as we would expect from Eq.~(\ref{eqz3Dcircular}) since the amplitude associated with the frequency $2\,n_{b}-3\,n$ is proportional to $(1+\cos{i})^2$. 

The next four examples are coplanar systems ($i=0$) with $e=0.1$ and $e_{b}=$0.1,~0.2,~0.3, and~0.4, respectively. When $e_{b}=0.1$ (Ex.~4), we detect the signals with frequencies  $2\,n_{b}-3\,n$ and $n_{b}-3\,n$ (cf.~Fig.~4 with $A=4.46$). When $e_{b} \ge 0.2$ (Exs.~5, 6 and 7), the signal with frequency $n_{b}-4\,n$ is above the detection limit (cf.~Fig.~4 with $A=4.46$) but we cannot distinguish it from the signal with frequency $n_{b}-3\,n$ since the resolution is not high enough. To resolve a frequency difference of $n$ we would indeed require a resolution of at least  $\Delta\,f=2\pi/n$, which corresponds to a time span  $T=2\pi/n$, i.e., about 22 years. At present, such a large timespan is not realistic since exoplanet detections started only 15 years ago. This lack of resolution prevents the correct detection of the signal with frequency $n_{b}-3\,n$ in the 5th, 6th, and 7th examples since this is "contaminated" by the close signal with frequency $n_{b}-4\,n$.  

In examples 1, 2, and 3, the binary mimics a planet while in examples 4, 5, 6, and 7, it mimics two planets with orbital periods in the ratio 2/1. In these cases, the data's precision and observation timespan do not allow us to distinguish the binary's effect from planet(s).  We now present an example where we detect signals with very close frequencies thus can  reject the planet(s) hypothesis.

Example 8 is a coplanar system ($i=0$) with $e_{b}=0.05$, and $e=0.4$.
According to Fig.~5, when $e=0.4$ the terms with frequencies $2\,n_{b}-6\,n$, $2\,n_{b}-5\,n$, $2\,n_{b}-7\,n$,
$2\,n_{b}-8\,n$, and $2\,n_{b}-9\,n$ are all above the detection level ($A=4.46$).
However, we only detect signals at approximately $2\,n_{b}-6\,n$ and $2\,n_{b}-8\,n$ as the observation timespan,
$t_{obs}\approx T/2$, is not long enough to resolve the other frequencies. These nearby  unresolved frequencies prevent the correct detection
of the signals at $2\,n_{b}-6\,n$ and $2\,n_{b}-8\,n$. Additional Fourier analysis of the residuals show the presence of a signal at $2\,n_{b}-10\,n$ but the fitting to the signals at $2\,n_{b}-6\,n$ and $2\,n_{b}-8\,n$ deteriorates when including this frequency. The residuals are 1.7 m/s, i.e., above the precision limit ($\sim 0.5$ m/s), which indicates the presence of unresolved / undetected frequencies. 

Finally, we note what happens when $e$ is large but the observation time span is short. We performed a new analysis of example 8 assuming $t_{obs}=2000$ days and $t_{obs}=1000$ days, i.e., about $T/4$ and $T/8$, respectively. In the first case, we detected signals at 243 days ($2\,n_{b}-6\,n$) and 281 days ($2\,n_{b}-11\,n$), with amplitudes 5.0 m/s and 2.7 m/s, respectively. However, the residuals after the fit are still 1.3 m/s. In the second case, we detect only one signal at 257 days ($\sim 2\,n_b$) with an amplitude 6.7 m/s,, which corresponds to a $0.3\,M_{J}$ planet at $0.79$ AU. Since $t_{obs}\ll T$, we cannot resolve the signals at nearby frequencies although these affect the measured amplitude. The residuals after the fit are $\sim 0.4$ m/s, hence there is no hint of unresolved / undetected frequencies.

\begin{table*}
\caption{Parameters of planets mimicked by several triple system configurations when
the instrument's precision is 0.535 m/s and the timespan is 4000 days. Radial velocity Fourier spectrum peak has amplitude $K_p$ and frequency $n_p$. Fake associated planet has period $T_{p}=2\pi/n_{p}$, semi-major axis $a_p$, mass $M_p$, and eccentricity $e_p$.} 
\label{table1}      
\begin{tabular}{|c |c |c c c | c c c c c|}     
\hline
&              & \multicolumn{3}{c}{Theory} & \multicolumn{5}{c|}{Simulations} \\ \hline
Ex. & Parameters    & $n_p$ & $T_{p}$ (d) & $K_p$ (m/s)  & $T_{p}$ (d) & $K_p$ (m/s)  & $a_p$ (AU) & $M_p$ & $e_p$  \\
\hline \hline
1 & $i=0^\circ$ & $n$ & 8069 & - & $8032.8\pm 2.6$ & $6922.0\pm 0.3$ & 9.97 & $1.05\,M_{\odot}$ & $0.00$ \\
  & $e=e_{b}=0$ & $2\,n_{b}-3\,n$ & 222.6 &  0.961 & $223.2\pm 0.5$  & $0.96\pm 0.06$ & 0.720 & $18.7\,M_{\oplus}$ & $0.00$  \\ \hline

2 & $i=30^\circ$ & $n$ & 8069 & - & $8037.6\pm 2.7$ & $6916.1\pm 0.3$ &  9.97 & $1.05\,M_{\odot}$ & $ 0.00$ \\
  & $e=e_{b}=0$ & $2\,n_{b}-3\,n$ & 222.6 & 0.837 & $222.9\pm 0.6$ & $0.83\pm 0.06$ &  0.719 & $16.0\,M_{\oplus}$ & $0.00$ \\ \hline

3 & $i=60^\circ$ & $n$ & 8069 & - & $8048.3\pm 2.7$ & $6904.5\pm 0.3$ &  9.98 & $1.05\,M_{\odot}$ & $0.00$ \\
  &$e=e_{b}=0$ & $2\,n_{b}-3\,n$ & 222.6 &  0.541 & $222.3\pm 0.9$ &  $0.52\pm 0.06$ & 0.718 & $10.16\,M_{\oplus}$ & $0.00$ \\ \hline

4 & $i=0^\circ$ & $n$ & 8069 & - & $8035.1\pm 2.8$ & $6958.1\pm 0.4$ &  9.97 & $1.05\,M_{\odot}$ & $0.10$ \\
  & $e=0.1$ & $2\,n_{b}-3\,n$ & 222.6 & 0.869 & $222.6\pm 0.6$ & $0.90\pm 0.06$ & 0.719 & $17.52\,M_{\oplus}$ & $0.00$ \\ 
  & $e_{b}=0.1$ & $n_{b}-3\,n$ & 485.5 & 0.585 & $485.4\pm 3.7$ & $0.65\pm 0.07$ &  1.209  & $16.01\,M_{\oplus}$ & $0.00$ \\ \hline

5 & $i=0^\circ$ & $n$ & 8069 & - &  $8026.8\pm 2.8$ & $6959.4\pm 0.4$ &  9.97 & $1.05\,M_{\odot}$ & $0.10$ \\
  & $e=0.1$ & $n_{b}-3\,n$ & 485.5 & 1.149 &  $485.2\pm 1.8$ & $1.27\pm 0.07$ & 1.208 & $31.9\,M_{\oplus}$ & $0.00$ \\
  & $e_{b}=0.2$ & $2\,n_{b}-3\,n$ & 222.6  &  0.804 &  $222.2\pm  0.6$ & $0.84\pm 0.06$ &  0.718  & $16.2\,M_{\oplus}$ & $0.00$ \\ \hline

6 & $i=0^\circ$ & $n$ & 8069 & - & $8074\pm 6$ & $6993\pm 5$ &  10.02 & $1.06\,M_{\odot}$ & $0.10$ \\
  & $e=0.1$ & $n_{b}-3\,n$ & 485.5 & 1.673 &  $476.8\pm 1.4$ & $1.53\pm 0.07$ & 1.194 & $38.5\,M_{\oplus}$ & $0.00$ \\
  & $e_{b}=0.3$ & $2\,n_{b}-3\,n$ & 222.6 & 0.701 &  $222.6\pm 0.8$ & $0.68\pm 0.06$ & 0.719  & $13.36\,M_{\oplus}$ & $0.00$ \\ \hline

7 & $i=0^\circ$ &  $n$ & 8069 & - & $8103\pm 6$ & $7020\pm 6$ & 10.06 & $1.07\,M_{\odot}$ & $ 0.10$ \\
  & $e=0.1$ & $n_{b}-3\,n$ & 485.5 & 2.135 &  $477.6\pm 1.1$ & $1.98\pm 0.07$ &  1.196 & $50.1\,M_{\oplus}$ & $ 0.00$ \\
  & $e_{b}=0.4$ & $2\,n_{b}-3\,n$ & 222.6 &  0.569 &  $223.1\pm 0.9$ & $0.57\pm 0.06$ &  0.720  & $11.2\,M_{\oplus}$ & $0.00$ \\ \hline

8 & $i=0^\circ$ &  $n$ & 8069 & - & $7928\pm 2$ & $7550.1\pm 0.4$ & 9.88 & $1.05\,M_{\odot}$ & $0.39$ \\
  & $e=0.4$ & $2\,n_{b}-6\,n$ & 242.7 & 1.104 &  $241.1\pm 0.3$ & $2.60\pm 0.07$ &  0.758 & $51.4\,M_{\oplus}$ & $ 0.12$ \\
  & $e_{b}=0.05$ & $2\,n_{b}-8\,n$ & 258.2 &  0.840 &  $253.7\pm 0.4$ & $2.52 \pm 0.07$ &  0.784  & $51.1\,M_{\oplus}$ & $0.00$ \\ 

\hline
\end{tabular}
\end{table*}

\begin{table*}
\caption{Triple system of solar mass star and binary on circular coplanar orbits: parameters of planets mimicked by the binary.}
\begin{tabular}{|c| c c c c c | c c c c |}
\hline
Ex. & $M_1$ ($M_{\odot}$) & $M_2$ ($M_{\odot}$) & $a$ (AU)& $a_b$ (AU) & $T$ (y) & $K_p$ (m/s) & $T_p$ (y) & $a_p$ (AU) & $M_p$ ($M_{\oplus}$) \\ \hline
1 & 0.3 & 0.3 & 10 & 1.0 & $10.26$ & $0.541$ & $0.685$ & $0.777$ & $16.00$ \\
2 & 1.0 & 1.0 & 10 & 1.0 & $10.26$ & $0.987$ & $0.375$ & $0.520$ & $23.90$ \\
3 & 1.0 & 1.0 & 50 & 5.0 & $114.7$ & $0.447$ & $4.197$ & $2.60$ & $23.90$ \\
4 & 1.0 & 1.0 & 50 & 2.5 & $114.7$ & $0.039$ & $1.482$ & $1.30$ & $1.494$ \\
5 & 1.0 & 1.0 & 100 & 10.0 & $324.4$ & $0.312$ & $11.86$ & $5.20$ & $23.90$ \\
6 & 1.0 & 1.0 & 100 & 5.0 & $324.4$ & $0.028$ & $4.197$ & $2.6$ & $1.494$ \\ \hline
\end{tabular}
\end{table*}

\subsection{Scaling of the observables}

In the previous section, we presented examples of several configurations of a triple star system where the
binary is at $a=10$~AU from the target star.
However, if the binary's stars are not both faint, it may be difficult to obtain precise radial velocity measurements 
because the target star's spectrum is likely to be contaminated by light received from the binary's stars. This will cause spectral line blending and it may be very difficult to separate the target star's motion from the binary's motion. Therefore, a realistic scenario would be a distant binary system (e.g.~50 or 100 AU) or a close binary (e.g.~10 AU)
composed of faint stars (e.g.~M-type).

In Table 2, we show some examples of triple systems composed of a solar mass star and a binary system on circular coplanar orbits. In examples~1 and 2, the planets could be detected by the current most precise instruments (e.g.~HARPS,~\citet{Harps2003}). On the other hand, the effect of the binaries in examples~3 and 5 is slightly below the current detection limit but the long periods imply that we would need a long observation timespan to constraint the orbits.  Finally, the planets mimicked by the binaries in examples~4 and 6 are not detectable by the current techniques but could be accessible to the planned exoplanet search program EXPRESS0/CODEX \citep{EXPRESSOCODEX2007}. 

Our results can be easily scaled: 
\begin{itemize}
\item If we increase $a$ and $a_b$ by a factor $\gamma$, then the ratio $a_{b}/a$ does not change, $K_p$ decreases by $\gamma^{-1/2}$ (Eq.~\ref{amplitudek}), $a_p$ increases by a factor $\gamma$ (Eq.~\ref{semajoraxis}), and  $M_p$ remains unchanged (Eq.~\ref{minmasspl}), while the periods increase by $\gamma^{3/2}$. 
\item If we decrease $M_1$ and $M_2$ by a factor $\gamma$, then $K_p$ decreases by $\gamma^{1/2}$ (Eq.~\ref{amplitudek}), $T_p$ increases by $\gamma^{-1/2}$, $a_p$ increases by a factor $\gamma^{-1/3}$ (Eq.~\ref{semajoraxis}), and  $M_p$ decreases by $\gamma^{1/3}$  (Eq.~\ref{minmasspl}), while $T$ remains unchanged.  
\end{itemize}

\subsection{The triple system HD~188753}

The system HD~188753 is composed of a star (A) and a close binary (Ba+Bb) with an orbital period of $155$ days 
that orbits the star (A) with a $25.7$ year period \citep{Konacki2005}.
A hot Jupiter  with a $3.35$ day period orbit around the A star was announced by \citet{Konacki2005} but this was later challenged by \citet{Eggenberger_etal2008} who claimed they could not find evidence for this planet. This shows the difficulty in identifying planets of stars with a nearby binary system. As noted previously, when observing the target star to obtain its spectrum we are likely to also detect light from the nearby stars. In the case of HD~188753, the contribution of Bb is modest since this star is faint but we cannot ignore the contribution of Ba, which is blended within the target star A's spectrum \citep{Eggenberger_etal2008}. 

The parameters of this triple system are $M_{\star}=1.06\,M_{\odot}$, $M_{1}=0.96\,M_{\odot}$, $M_{2}=0.67\,M_{\odot}$,   $a=12.3$~AU, $a_{b}=0.67$~AU, $e=0.5$, $e_{b}=0.1$, and $I=34^\circ$ \citep{Konacki2005}. We can apply our theory to estimate the effect of the binary on the star assuming $50$ data points and $t_{obs}=500$ days, which are approximately the parameters reported in \citet{Eggenberger_etal2008}. Assuming that the triple system is coplanar, we obtain radial velocity oscillations with an 85 day period (frequency $\sim 2\,n_b$) and amplitude 0.5 m/s (i.e., 5.5 times the value obtained with a circular coplanar model). This can be interpreted as a fake planet of 4 Earth masses located at 0.38 AU.
As $t_{obs} \ll T=25.7$ years, we cannot resolve additional frequencies near $2\,n_{b}$. However, the predictions made here will change for different values of the angle variables (phases and relative inclination) and if we increase the observation time span.  
Nevertheless, the precision obtained for this system is  currently only $60$ m/s \citep{Eggenberger_etal2008}, hence this effect is not detectable.

\subsection{Planets in binaries}

In \citet{Morais&Correia2008}, we reviewed all currently known planets of stars which are part of a moderately close binary system ($a\la 100$ AU). Our aim was to see whether these could be fake planets due to a binary composed of the companion star, $M_1$, and another unresolved star, $M_2$. We computed the parameters of the hidden binary component ($M_2$ and $a_b$) that could mimic these planets (Table 1, \citet{Morais&Correia2008}) and saw that $M_2$ would either be too massive to be realistic or at least too massive to be unreported. 

The results in \citet{Morais&Correia2008} are only valid in the context of triple star systems on circular coplanar orbits. Here, we saw that when the triple system's orbits are eccentric, the magnitude of the binary's effect can increase with respect to the circular case (Secs.~4.1 and 4.3). Therefore, in the eccentric case a planet can be mimicked by a less massive binary than in the circular case.

The most interesting cases in Table 1 \citep{Morais&Correia2008}, namely $\gamma$\,Cep and HD\,196885, are binary systems with eccentric orbits. 
The star $\gamma$\,Cep~A orbits $\gamma$\,Cep~B with $e=0.41$ and $I=119^\circ$ \citep{Neuhaeuser_etal_2007, Mugrauer_etal_2008}, while $t_{obs}\approx T/2$ \citep{Torres_2007}. 
The star HD\,196885~A orbits HD\,196885~B with $e=0.46$ but $I$ is unknown while $t_{obs}\approx T/4$ \citep{Correia_etal_2008}.
However, extrapolating from the results in Sect.~4.1, if there were hidden companion stars close to $\gamma$\,Cep~B or HD\,196885~B,  we would expect to be able to detect, in each case, several nearby peaks and not single peaks at the planets' orbital frequencies.

Moreover, the stars $\gamma$\,Cep~A+B have been directly imaged and  $\gamma$\,Cep~B's brightness is consistent with that of a single M4-type dwarf  \citep{Neuhaeuser_etal_2007}. 
Additionally, we performed some 3-body fits to the data using a model with HD\,196885~A+B and an unresolved star nearby  HD\,196885~B. We saw that these fits were worse (residuals at least 31 m/s and poorly constrained orbital elements) than fits using a model with the binary star system 
(HD\,196885~A+B) and a planet companion to HD\,196885~A (residuals of 11 m/s in agreement with \citet{Correia_etal_2008}). Therefore, the planets around $\gamma$\,Cep~A and  HD\,196885~A, respectively, are not likely to be due to hidden star companions to $\gamma$\,Cep~B or HD\,196885~B.

\section{Conclusion}

We have studied a triple system composed of a star and a binary system. The star and binary system have eccentric and inclined orbits. This is an extension of earlier work where we assumed that the star and binary system are on circular coplanar orbits \citep{Morais&Correia2008}.

We demonstrated that if we are unaware of the binary system’s presence (one or even both components may be unresolved for instance because they are faint M stars) we may then be led to believe that the star has one or even two planet companions.
Although the radial velocity variations due to a binary are distinct from those due to planet(s), in practice, the measured effect depends on the instrument's precision and the observation time span.

We have shown that because of the limited instrumental precision, we may only detect periodic terms with well separated frequencies hence we mistake these for planet(s).
We also saw that, if the observation time span is shorter than the star's long-period motion around the binary's center of mass, we may not be able to resolve terms with nearby frequencies, which means that we cannot distinguish fake planet(s) from  a binary. 

We have also demonstrated that the binary's effect is more likely to be mistaken for planet(s) when the  radial velocity oscillations are composed of large dominant periodic terms with well separated frequencies. This is more likely to happen in the case of coplanar orbits observed equator-on. Moreover, we have seen that when the orbits are eccentric, the magnitude of the binary's effect can increase with respect to the circular case. Nevertheless, our model is valid for any triple system's configuration.

We have presented an example of a binary system that affects a nearby star's motion. When the star's long-period motion 
has low to medium eccentricity, the binary can mimic planet(s) of 10 to 50 Earth masses, which are near the current detection limit. However, if the long-period motion has a high eccentricity we are more likely to detect multiple signals with very close frequencies and therefore reject the planet hypothesis. An exception occurs when the long period motion has a high eccentricity but its period is poorly constrained (due to a short observational time span) in which case we may mistake the binary for a planet of about 100 Earth masses.

We showed that when the binary has an eccentric orbit it can mimic two planets with periods approximately in the ratio 2/1. Therefore we may be misled to think that we found planets in the 2/1 orbital resonance when we have an eccentric hidden binary. This is somehow analogous to the case studied by \citet{Anglada-Escude2010} where two planets on circular orbits in the 2/1 orbital resonance can be mistaken for a single planet with an eccentric orbit. However, our scenario is probably more realistic since planets in orbital resonances are not likely to have circular orbits.

We propose that new planet detections in close binary systems, especially Earth-sized objects that are the targets of the planned search program EXPRESSO/CODEX \citep{EXPRESSOCODEX2007}, be checked carefully because they could indeed be artifacts caused by a hidden binary. 
This could be done by comparing fits to the data using (1) a model composed of the binary star system and a planet with (2) those obtained for a model composed of a hierarchical triple star system. If the fits of (2) are at least as good as (1), then the hidden binary hypothesis should be considered.

\begin{acknowledgements}
We acknowledge financial support from FCT-Portugal (grant PTDC/CTE-AST/098528/2008).
\end{acknowledgements}

\bibliographystyle{aa}

\end{document}